\documentclass[
 reprint,=
 amsmath,amssymb,
 aps,unsortedaddress]{revtex4-1}
\usepackage{xcolor}
\usepackage{graphicx}
\usepackage{dcolumn}
\usepackage{bm}
\usepackage{amsmath}
\usepackage{lipsum}
\usepackage{mathtools}

\newcommand{\be}{\begin{equation}}
\newcommand{\ee}{\end{equation}}

\begin{document}

\title{Higher-order topological phase without crystalline symmetry}

\author{Yizhi You}
\affiliation{\mbox{Princeton Center for Theoretical Science, Princeton University, Princeton NJ 08544, USA}}

\date{\today}
\begin{abstract}
A wide variety of higher-order symmetry protected topological phase(HOSPT) with gapless corners or hinges had been proposed as a descendant of topological crystalline insulator protected by spatial symmetry. 
In this work, we address a new class of higher-order topological state which does not require crystalline symmetries but instead relies on subsystem symmetry for protection.
We propose several strong interacting models with gapless hinge or corner based on a `decorated hinge-wall condensate'  picture.  The hinge-wall, which appears as the defect configuration of $Z_2$ paramagnet is decorated with lower-dim SPT state. Such unique hinge-wall decoration structure leads to gapped surfaces separated by gapless hinges. The non-trivial nature of the hinge modes can be captured by a $1+1$D conformal field theory with a Wess-Zumino-Witten term. Besides, we establish a no-go theorem to demonstrate the ungappable nature of the hinge by making a connection between generalized Lieb-Schultz-Mattis theorem and the boundary anomaly of HOSPT state. This universal correspondence engenders a comprehensive criterion to determine the existence of HOSPT under certain symmetry regardless of the microscopic Hamiltonian.
\end{abstract}

\maketitle

\section{Introduction}
Quantum many-body system prompts exotic phases of matter enriched by strong interactions and quantum entanglement.
Tremendous effort had been made toward the classification and characterization of topological materials along with their bosonic descendants in the presence of internal and crystalline symmetries \cite{fu2011topological,hsieh2012topological,cheng2016translational,ando2015topological,slager2013space,hong2017topological,qi2015anomalous,huang2017building,teo2013existence,song2017topological,watanabe2017structure,po2017symmetry,isobe2015theory}. In addition to fully dispersive boundary modes, topological crystalline phases admit gapped edges or surfaces with protected gapless modes at high-symmetry corners or hinges. Exemplifying a much richer bulk-boundary correspondence, this phenomenology is now termed higher-order symmetry protected topological phase(HOSPT)~\cite{benalcazar2017quantized,benalcazar2017electric,langbehn2017reflection,song2017d,song2017d,dwivedi2018majorana}. Beyond the non-interacting higher-order topological insulator and superconductor triggered by non-trivial band topology, there appears a variety of boson (or interacting fermion) candidates for HOSPT with gapless corners or anomalous hinge states\cite{tiwari2019unhinging,you2018majorana,you2018higher2}. 
Apart from the mathematical characterization, the physical manifestations and material fabrications of HOSPT have appeared in a variety of platforms\cite{peterson2018quantized,schindler2018higher,serra2018}. Moreover, there is increasingly compelling evidence to show that some higher-order topological superconductor corners can support exotic fractionalized quasi-particles including parafermion or projective Ising anyon\cite{you2018higher,laubscher2019fractional}, providing new architectures for quantum information processing and quantum computation.

At this stage, all the prominent examples of higher-order topology require crystalline symmetries. Due to the restriction of crystalline symmetry, the gapless corner or hinge modes are distributed in a spatially symmetric way. In the absence of spatial symmetry, one can typically hybridize and remove the spatially separated corner(hinge) modes through an edge(surface) phase transition without the bulk gap closing. It remains unclear whether there exists a HOSPT state beyond crystalline symmetry protection\footnote{Yizhi You, Taylor L. Hughes, F.~J. Burnell, To appear}. 

In this work, we address a new class of HOSPT phase without crystalline symmetries, but instead relies on subsystem symmetry for protection.
In $D$ spatial dimensions, a subsystem symmetry consists of independent symmetry operations acting on a set of d-dimensional subsystems with $0<d<D$. In $D=3$, the subsystems can be lines ($d=1$), planes($d=1$) or fractals \cite{Xu2004-oj,you2018subsystem,devakul2018fractal,Vijay2016-dr,you2018symmetric}. The corresponding subsystem symmetry generates a quantum number that is conserved separately on each sub-manifold, leading to interesting new possibilities for both symmetry-breaking \cite{Xu2004-oj} and symmetry-protected topological phases\cite{you2018subsystem,devakul2018fractal,devakul2018strong,shirley2018foliated,shirley2018fractional,shirley2018foliated,shirley2019twisted}. In Ref.~\cite{you2018subsystem,devakul2018fractal,devakul2018strong,shirley2018foliated,shirley2019twisted}, the authors introduced a zoology of subsystem protected topological phase with gapless boundaries. Here, we propose a parallel 3D subsystem symmetric higher-order topological phase which supports gappable surfaces but gapless corners or hinges. Due to the subsystem symmetry which induces charge conservation on each 2D plane, the spatially separated gapless hinge or corner modes are robust against any symmetry allowed perturbations. It is worth emphasizing that the subsystem symmetric HOSPT is an interaction enabled higher-order topological phase.
These nontrivial phases require putative strong interactions for their existence, and in the cases we consider, the free-fermion classification yields only a trivial phase. 

To be more explicit, the general construction of subsystem symmetric HOSPT state depends on a decorated hinge-wall picture. In Ref.~\cite{huang2018cage,prem2017emergent,song2018twisted,you2019fractonic,shirley2018fractional,slagle2019foliated,williamson2016fractal,Williamson2016-lv}, it is demonstrated that gauging a subsystem symmetry leads to Fracton topological order with cage-net(or membrane-cage-net) structure. By decorating such membrane-cage-net structure with a nontrivial lower-dim SPT state, the hinge(corner) carries fractional quantum number or symmetry anomaly and thus engenders a symmetry enforced gapless hinge (corner). Such construction, which we will present in the rest of the paper is illuminating because it allows for a direct field-theoretic connection between subsystem HOSPT and symmetry enriched Fracton gauge theories\cite{you2019fractonic,shirley2019universal}.

Besides subsystem symmetric HOSPT state from exactly solvable models, we also propose a general criterion for the existence of some particular subsystem symmetric HOSPT phases by making a connection between generalized Lieb-Schultz-Mattis(LSM) theorem and the boundary anomaly of a 3D HOSPT phase. Precisely, for a 3D HOSPT state protected by subsystem symmetry($G^{sub}$) and global on-site symmetry($S$), its symmetric boundary theory can be mapped into a 2D lattice model with subsystem symmetry($G^{sub}$). In addition, the lattice symmetry of such 2D system acts in a similar way as the on-site symmetry($S$) at the boundary of HOSPT. Such mapping between onsite symmetry at the boundary of 3D HOSPT and lattice symmetry in 2D lattice models allows us to determine the possibility of a featureless gapped boundary. In particular, for 2D lattice models with subsystem symmetry\footnote{Huan He, Yizhi You, Abhinav Prem, To appear}, there is a general Lieb-Schultz-Mattis Theorem\cite{oshikawa2000commensurability,hastings2004lieb} excluding the possibility of a featureless gapped phase hosting a unique ground state compatible with lattice and subsystem symmetry. The absence of such featureless gapped phase in 2D implies the 3D HOSPT boundary, with an onsite symmetry $S$ playing a role akin to the 2D lattice symmetry, does not admit a trivially gapped boundary(including the hinge) in the presence of onsite symmetry $S$ and subsystem symmetry $G^{sub}$. Subsequently, such ungappable boundary with anomalous symmetry must be accompanied by a nontrivial bulk with higher order topology.
Based on such well-established Lieb-Schultz-Mattis Theorem \cite{else2019topological,jiang2019generalized,cheng2019fermionic,jian2018lieb,bultinck2018filling,hsieh2016all,thorngren2018gauging}, one can readily determine the existence of HOSPT via its boundary actions even in the absence of any concrete Hamiltonian. As the Lieb-Schultz-Mattis theorem is universal for any strongly interacting systems, the corresponding ungappable condition for HOSPT boundary always applies regardless of the microscopic Hamiltonian.

\section{3D HOSPT with protected corner mode}

To start with, we first reboot our brain by looking into a simple exactly solvable model on a cubic lattice. This model displays a HOSPT with Majorana corner mode protected by subsystem fermion parity symmetry. Since our main tool is to study exactly solvable model Hamiltonians, fermions introduce a new technical challenge: a fermion parity subsystem symmetry requires interactions that are at least quartic in the fermion operators.  
The resulting Hamiltonians are generally not solvable unless the interaction terms treat non-overlapping sets of fermions -- and necessarily not in the same symmetry class as any non-interacting topological phases of fermions. 
This also implies the HOSPT protected by subsystem symmetry is unique in strongly interacting systems without any band-fermion analogy. Here we will give one example, building a 3D charge 4e superconductivity with subsystem fermion parity symmetry on each $i-j$ plane.

\begin{figure}[h]
  \centering
      \includegraphics[width=0.5\textwidth]{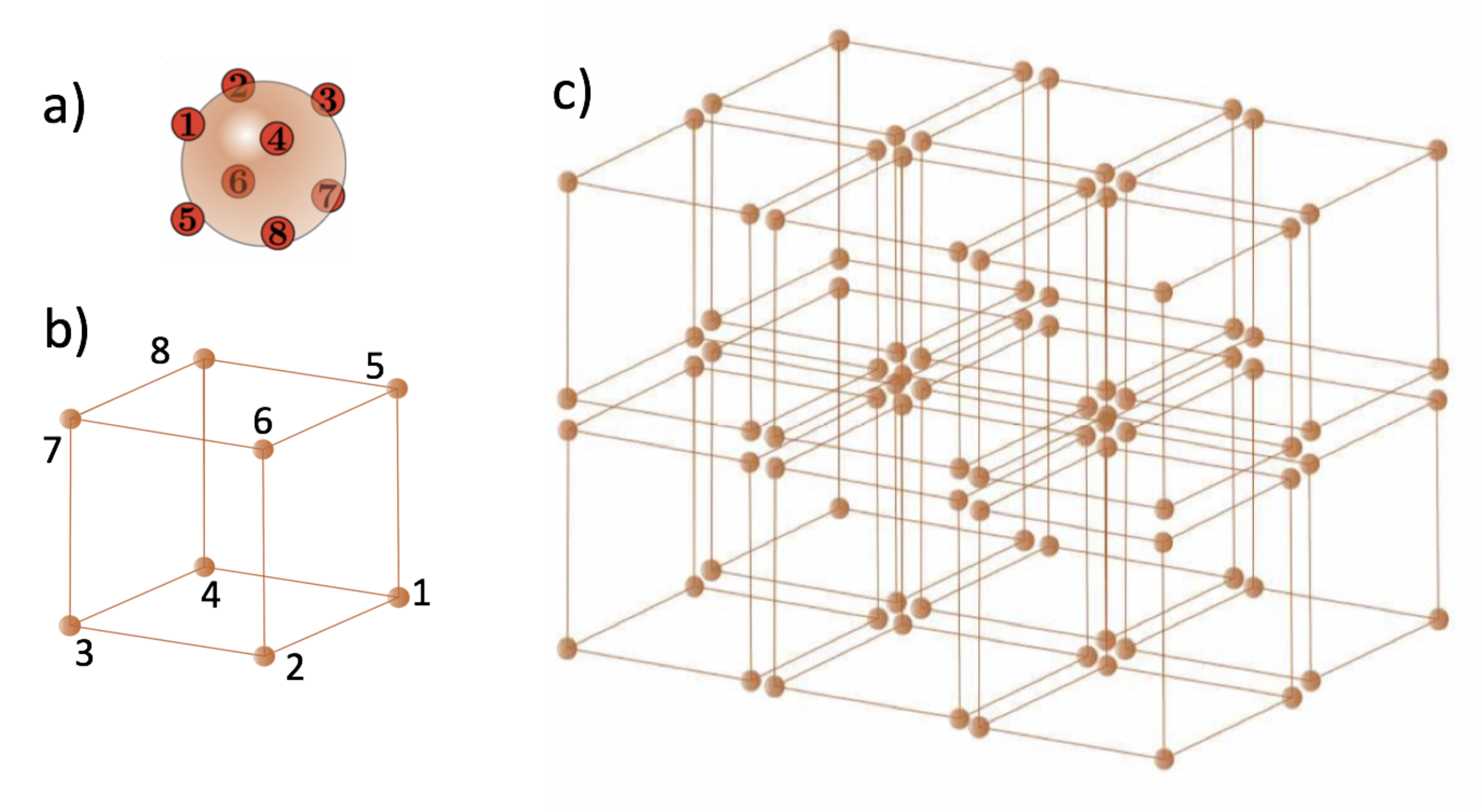}
  \caption{a) Each site has eight Majoranas $\eta_1,..\eta_8$. b-c) The eight Majoranas living at the corner of each cube is projected into a unique ground state preserving coplanar fermion parity symmetry. } 
  \label{fermion}
\end{figure}

We begin with 4 fermions on each site of the cubic lattice as Fig.~[\ref{fermion}]. One can decompose the four fermion on each site into eight Majorana operators labelled as $\eta_1,..\eta_8$.
On each cube, there are several Majorana quartet interactions among the eight Majoranas living at the cube corners. Of the eight fermions on each site, each participates in the cluster interaction term in one of the eight nearby cubes, so that all interactions commute.

 To describe the interaction terms, we label the eight Majorana as Fig.~[\ref{fermion}]. 
 Each cube cluster containing 8 Majorana fermions are coupled via Fidkowski-Kitaev\cite{fidkowski2010effects,fidkowski2011topological} type interactions.
 Specifically, 
we first add a 4-Majorana interaction among the top and bottom surface as
\begin{align} 
H_1=\eta_5\eta_6\eta_7\eta_8+\eta_1\eta_2\eta_3\eta_4  .
\label{four}
\end{align}
Ground states of $H_1$ can be described via the complex fermions,
\begin{align} 
&\Psi_{\uparrow}=\eta_5+i\eta_6,\Psi_{\downarrow}=\eta_7+i\eta_8\nonumber\\
&\Psi'_{\uparrow}=\eta_1+i\eta_2,\Psi'_{\downarrow}=\eta_3+i\eta_4
\label{mapf}
\end{align}
In these variables, the Hamiltonian $H_1$ becomes,
\begin{align} 
H_1=(n_{\Psi}-1)^2+(n_{\Psi'}-1)^2
\end{align}
Thus $H_1$ favors the odd fermion parity state for both $\Psi$ and $\Psi'$. 
This allow us to map the ground state subspace of $H_1$ into two spin $1/2$ degrees of freedom per cube:
\begin{align} 
&\vec{n}_i=\Psi^{\dagger} \vec{\sigma}_i \Psi, \vec{m}_i=\Psi'^{\dagger} \vec{\sigma}_i \Psi', 
\end{align}

In terms of these spin degrees of freedom, the second interaction on the cube cluster is,
\begin{align} 
& H_2=- m_x n_x-m_y n_y\nonumber\\
&= (\eta_5\eta_6-\eta_7\eta_8)(\eta_1\eta_2-\eta_3\eta_4)\nonumber\\
&+(\eta_5\eta_8-\eta_6\eta_7)(\eta_1\eta_4-\eta_2\eta_3)
\label{four2}
\end{align}
Such XY interaction  
projects the two spins in each cube cluster into an SU(2) singlet, yielding a unique ground state. With this cluster interaction on each cube in Fig.~[\ref{fermion}], the many-body Hamiltonian is fully gapped with a unique ground state in the bulk.

What are the symmetries of this model? Due to the quartet nature of the Majorana interaction, our Hamiltonian conserves the fermion parity of each $x-y,y-z,x-z$ plane separately. This subsystem fermion parity symmetry indicates any fermion-bilinear interaction between sites is prohibited so the leading order inter-site couplings are the quartet interactions.

In the presence of a boundary, each surface(or hinge) site contains four(or two) unpaired Majoranas which can be gapped out via onsite Majorana hybridization. When it comes to the corner, the corner site carries 7 free Majorana zero modes which cannot be gapped out via onsite hybridization. Besides, due to the subsystem fermion-parity symmetry, one cannot hybridize or gap out the Majorana zero mode from different corners via surface/hinge phase transitions(while keeping the bulk gap) as such coupling always breaks fermion parity on specific $i-j$ planes. Thus, the Majorana zero mode on each corner is robust under subsystem fermion-parity symmetry.

Several earlier literature\cite{benalcazar2018quantization,benalcazar2017electric,benalcazar2017quantized} propose a class of third-order topological insulator/superconductor in 3D with corner zero modes. However, these corner modes can be annihilated via hinge/surface phase transitions without bulk gap closing. Accordingly, these models are not intrinsically topological as the existence of corner mode is not connected to the bulk physics. To illustrate this subtlety we note that since there are \emph{three} hinges that terminate at a corner, then, in principle, one can always decorate the hinges with nontrivial 1D SPT chains on all hinges to cancel the corner modes. Subsequently, there is no `intrinsic fermionic topological octupole insulator' with robust corner modes protected by cubic symmetry\cite{rasmussen2018intrinsically,rasmussen2018classification}.
However, with the additional subsystem fermion parity symmetry, a fermionic HOSPT with gapless corner modes turns into a reality as the subsystem symmetry strictly constraints the interaction.
It is worth mentioning that our subsystem fermion parity symmetric HOSPT does not have a non-interacting counterpart, as any fermion hopping term breaks subsystem fermion parity in any case. Thus, the subsystem symmetric HOSPT is a unique feature in strongly interacting system. 

\section{Subsystem HOSPT with gapless hinges}

Thus far, we have built a 3rd-order topological phase with gapless corner modes
protected by subsystem symmetry. Specifically, such class of subsystem symmetric HOSPT always requires putative strong interaction without a free-fermion counterpart.
We will now proceed to formulate a second-order
topological phase with gapless hinges protected by subsystem symmetries. In particular, we are interested in the case where the many-body system carries both subsystem symmetry $G^{sub}$ and global symmetry $S$. If such state manifest a higher-order topological phase, the global symmetry defect $S$ at the hinge carries a fractional quantum number of $G^{sub}$ so the hinge becomes ungappable. In particular, the surface and hinge theory of such HOSPT can be mapped into a lower dimensional lattice system with the same subsystem symmetry($G^{sub}$). In the meantime, the lattice symmetry acts in a similar way as the on-site symmetry($S$) at the boundary of HOSPT. Such mapping between onsite symmetry at the boundary of 3D HOSPT and lattice symmetry in 2D lattice models allows us to determine the possibility of a featureless gapped boundary based on Lieb-Schultz-Mattis theorem. This noteworthy mapping engenders a general criterion for the existence of the higher-order topological phase under $G^{sub} \times S$ symmetry.

\subsection{Decorated hinge-wall model}
\begin{figure}[h]
  \centering
      \includegraphics[width=0.45\textwidth]{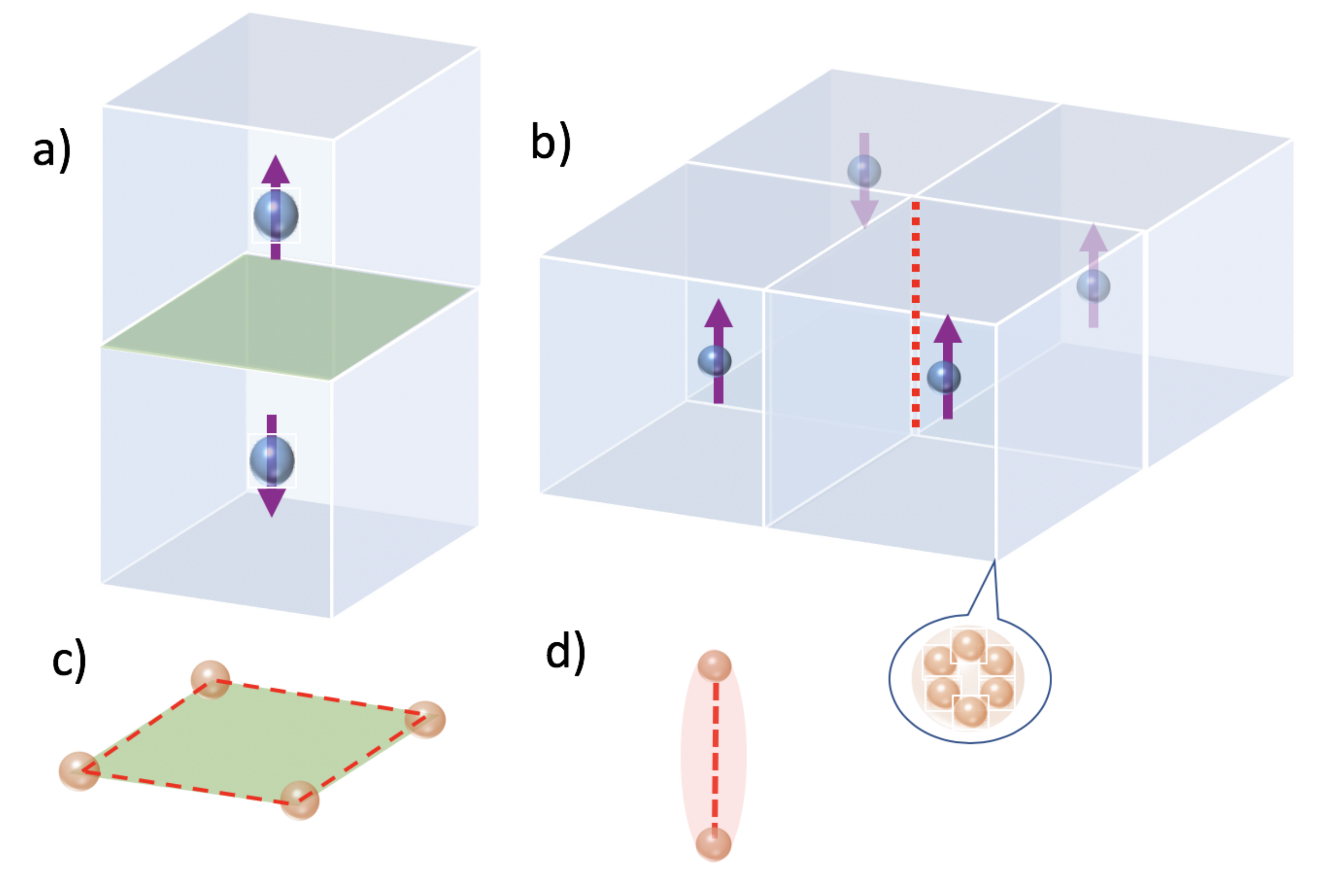}
  \caption{a) The $\sigma^z$ domain wall on the x-y  plane. b) The hinge defect along the z-link. c) Each x-y domain wall is decorated with a 2D HOSPT with protected spin 1/2 corner mode(red dot). d) Each z-hinge defect is deocrated with SU(2) spin singlet.} 
  \label{boson}
\end{figure}

To begin let us look into a spin model on BCC lattice in Fig.~\ref{boson}. The cube center contains a single Ising spin $\sigma$(blue dot) while the cube corner carries six spin-1/2 degree of freedom labelled as $\tau$(red dots).
The spin $\sigma$ living at the cube center is placed into a $Z_2$ paramagnetic phase,
\begin{align} 
&H_{\sigma}=-\sum_i \sigma^x_i
\end{align}
Such Hamiltonian has a global $Z_2$ symmetry generated by $\sigma^x$. In the paramagnetic phase, the $\sigma$ spin are polarized in the x-direction. If we choose the $\sigma^z$ basis, the ground state wave function is a coherent superposition of all domain wall membrane configurations on the cubic lattice which separate the regions between $\sigma^z=1$ and $\sigma^z=-1$. 

The interaction among the $\tau$ spins from the cube corner depends on the adjacent $\sigma$ spin structure. In particular, if the two $\sigma^z$ spins form a domain wall on the x-y plane as $\sigma^z(r+\frac{e_z}{2}) \sigma^z(r-\frac{e_z}{2})=-1$ in Fig.~\ref{boson}, we decorate such domain wall plaquette on the x-y plane by creating a four spin entangled pair $|\psi \rangle_{r+e_z}=|0101\rangle+|1010\rangle$ among the four $\tau$ spins on the corners of the domain wall plaquette. Such decoration can be expressed in terms of the projection Hamiltonian,
\begin{align} 
&H^1_{\sigma,\tau}= (1 +\sigma^z(r+\frac{e_z}{2}) \sigma^z(r-\frac{e_z}{2}))  |\psi_{r+e_z} \rangle  \langle \psi_{r+e_z}|
\label{p1}
\end{align}
$|\psi_{r+e_z}\rangle $ refers to the four spins entangled state on the corners of the domain wall plaquette. When the above and beneath $\sigma^z$ spins form a domain wall, four $\tau$ spins from the plaquette corner are projected into $\psi$ state. 
When it comes to the corner of the domain wall on x-y plane as Fig.\ref{boson}, the corner site contains odd number of unpaired spin-1/2 zero modes protected by $\mathcal{T}$. Meanwhile, all other sites contain even number of spin-1/2 which can be gapped out by pairing them into an onsite singlet. Based on this observation, each domain wall on x-y plane is embellished with a 2D HOSPT with protected spin-1/2 corner mode\cite{you2018higher}.

Apart from the $Z_2$ symmetry for $\sigma$, this Hamiltonian has a global time-reversal symmetry $\mathcal{T}$ for $\tau$ that flips each spin, and a subsystem U(1) symmetry which preserves the $\tau^z$ number on each xz and yz plane.
\begin{align} 
&\mathcal{T}= \mathcal{K} i\tau^y, \nonumber\\
& U^{sub}(1): \prod_{j \in P_{i-z}}e^{i \frac{\theta}{2} \tau^z_j}.
\label{sym}
\end{align}
 The subsystem U(1) symmetry acts only on the six $\tau$ spins inside each unit cell in each yz or xz plane, and rotates the $\tau$ spin around the $S_z$ axis. Hence, subsystem U(1) symmetry preserves the total $S_z$ charge of $\tau$ along any yz and xz plane, and forbids spin-bilinear coupling \emph{between} $S_x,S_y$ channel on x(or y) links. Furthermore, the global $\mathcal{T}$ symmetry forbids terms in the Hamiltonian that polarize the spins.

In addition, when the four $\sigma$ spins adjacent to each z-hinge contain odd number of $\sigma^z=-1$ as Fig.~\ref{boson}, we decorate the z-hinge by creating a two spin entangled pair $|\phi \rangle_{r}=|01\rangle-|10\rangle$ between the two $\tau$ spins at the end of z-link. Such decoration can be expressed in terms of the projection Hamiltonian,
\begin{widetext} 
\begin{align} 
&H^2_{\sigma,\tau}= (1+\sigma^z(r-\frac{e_x+e_y}{2}) \sigma^z(r-\frac{e_x-e_y}{2})\sigma^z(r+\frac{e_x+e_y}{2}) \sigma^z(r+\frac{e_x-e_y}{2}))~|\phi_{r} \rangle  \langle \phi_{r}|
\label{p2}
\end{align}
\end{widetext}
When the four $\sigma$ spins adjacent to the hinge have odd number of spin down state, the two $\tau$ spin between the hinge is projected into an SU(2) singlet. Such `hinge defect' can be regarded as the intersection line between two $\sigma$ domain walls from the yz and xz plane. The interaction in Eq.~\ref{p2} embellishes each `z-hinge defect' with an AKLT chain along the z-row. This AKLT chain decorated on the z-hinge preserves $U^{sub}(1)$ and global $\mathcal{T}$ symmetry defined in Eq.~\ref{sym}. As the interaction between $\tau$ spin only appears among the z-hinge, the $S_z$ quantum number on each xz and yz plane is still preserved. Since the z-hinge defect carries an AKLT chain, the open end of the z-hinge defect contains a spin-1/2 zero mode.

\begin{figure}[h]
  \centering
      \includegraphics[width=0.35\textwidth]{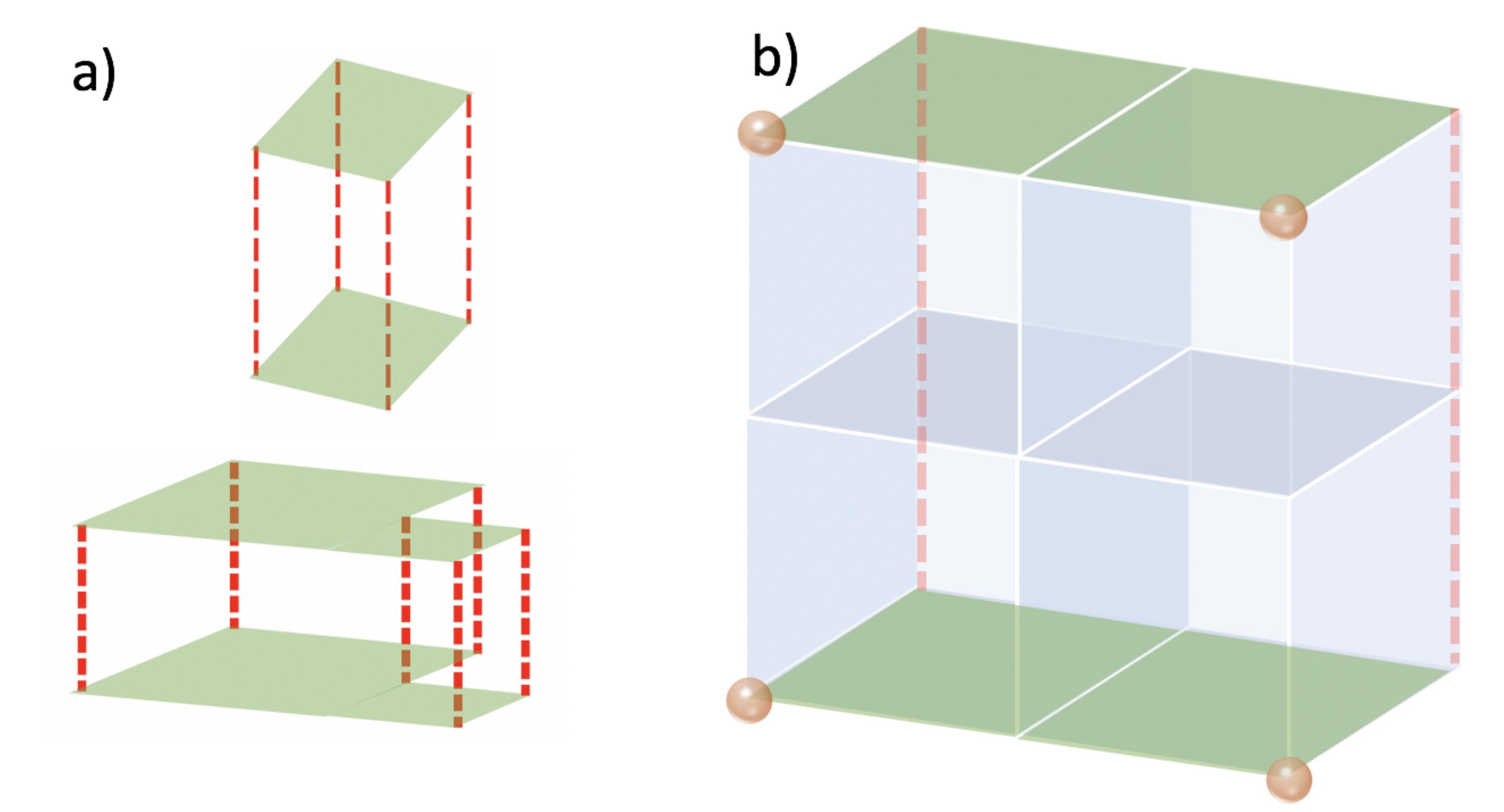}
  \caption{a) The ground state is a superposition of all possible domain wall configurations of $\sigma^z$. Here the green plaquette illustrates the domain wall on x-y plane, the red dashed line is the z-hinge as an intersection line between xz and yz plane domain walls. The domain wall on x-y plane is decorated with 2D HOSPT with spin 1/2 corner mode while the z-hinge carries and AKLT chain along z-direction. b) When the green domain wall on x-y plane hits the side boundary, the corner carries an unpaired spin-1/2.} 
  \label{boson2}
\end{figure}

The Hamiltonian we construct here has a simple ground state wave function. As the $\sigma$ spin living at the cube center is in the $Z_2$ paramagnetic phase, its ground state can be expressed in terms of superposition of all close domain wall configurations in the bulk. Meanwhile, due to the $\tau$ spin decoration in Eq.~\ref{p1}-\ref{p2}, the domain wall on each x-y plane contains a 2d HOSPT with spinon corner mode while the z-hinge defect is decorated with an AKLT chain. As the z-hinge defect is nothing but the domain wall intersection line between xz and yz plane, any open end of the z-hinge defect is connected to the domain wall corner on the xy plane as Fig.~\ref{boson2}.  Such connection is a consequence of close domain wall configuration in the bulk, and the connecting point between z-hinge end and x-y domain wall corner is merely the intersection point between three domain wall planes from xy, yz, xz directions. As a result, the ground state can be viewed as a hinge-wall condensate with each z-hinge connecting a domain wall corner from the xy plane.
The spin-1/2 zero mode living at the corner of x-y domain wall can hybridize with the dangling spin-1/2 living at the end of the z-hinge defect. Subsequently, the hinge-wall condensation give rise to a gapped state in the bulk.

When it comes to the surface, we first focus on the side faces on yz and xz planes. As is illustrated in Fig.~\ref{boson2}, the corner of the x-y domain wall hitting at the side faces on yz(or xz) plane is no longer connected by a z-hinge. Such open x-y domain wall corner at the boundary carries a spin-1/2 zero mode due to its 2D HOSPT decoration. We now construct a symmetry preserving surface perturbation which fully gap out the degree of freedom on the side faces.
\begin{align} 
&H_{\text{x-z surface}}= (1+\sigma^z(r-\frac{e_x}{2}) \sigma^z(r+\frac{e_x}{2}))~|\phi_{r} \rangle  \langle \phi_{r}|,\nonumber\\
&H_{\text{y-z surface}}= (1+\sigma^z(r-\frac{e_y}{2}) \sigma^z(r+\frac{e_y}{2}))~|\phi_{r} \rangle  \langle \phi_{r}|
\label{p3}
\end{align}

\begin{figure}[h]
  \centering
      \includegraphics[width=0.3\textwidth]{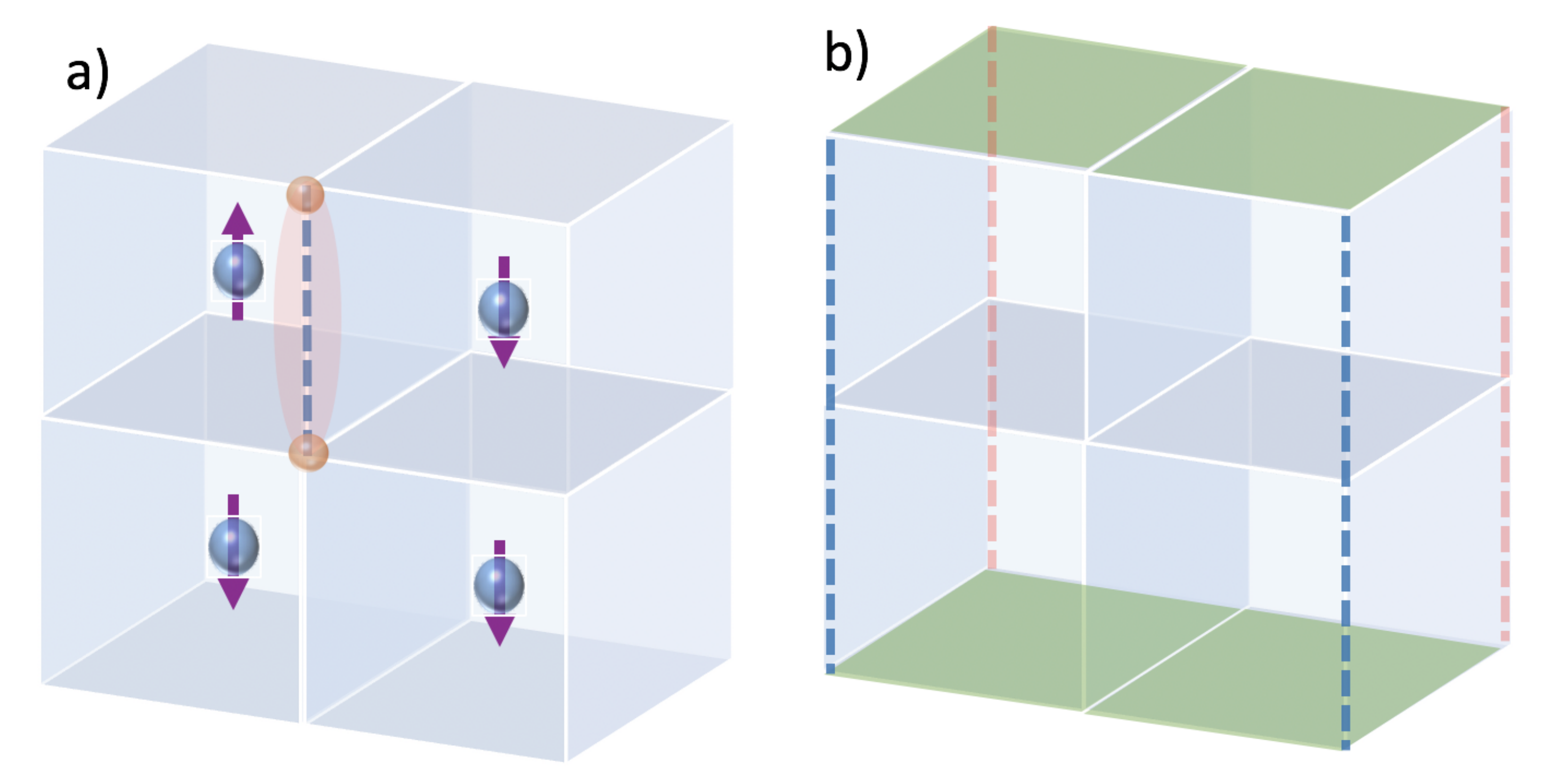}
  \caption{ a) The z-domain wall line on the side face between two anti-parallel $\sigma^z$ is decorated with a spin singlet formed by $\tau$. b) The x-y domain wall corner hitting at the side surface is connected to the surface domain wall line along z direction. As the  z-domain wall line on the side face is decorated with an AKLT chain, the connecting point can be fully gapped.} 
  \label{boson3}
\end{figure}

Such surface interaction decorates the z-directional domain wall line on each each xz(yz) surface with an AKLT chain formed by $\tau$ spin. As the AKLT chain is an inter-site dimer singlet between z-links, it still preserves the $U^{sub}(1)$ on xz and yz planes as well as global $\mathcal{T}$ symmetry. The surface domain wall line along z-direction are connected by the corner of x-y domain wall hitting the boundary as Fig.~\ref{boson3}. Such connecting point contains two spin-1/2 zero modes contributed by the x-y domain wall corner and the z-directional surface domain wall end. By coupling these two spin-1/2 zero modes, the surface area is fully gapped.

\begin{figure}[h]
  \centering
      \includegraphics[width=0.15\textwidth]{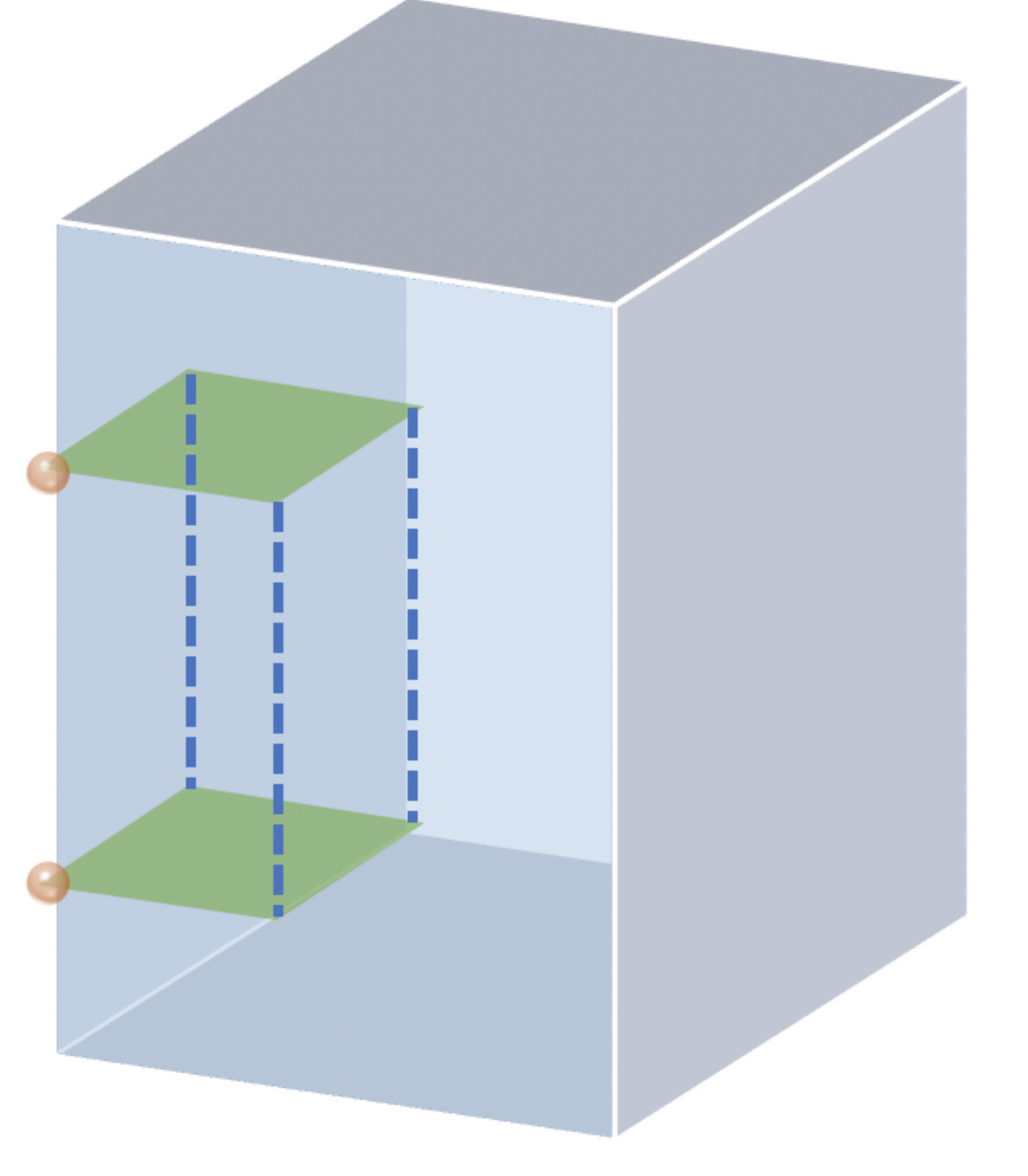}
  \caption{When the the corner of x-y domain wall hit the hinge, there appears a free spin-1/2 zero mode(red dot). The proliferation of such domain wall at the hinge carrying spinon mode leads to a gapless hinge state. } 
  \label{bosonhinge}
\end{figure}

We now consider the degrees of freedom on the z-hinge. Based on our bulk Hamiltonian, if there being a domain wall between $\sigma^z(r-\frac{e_z}{2}),\sigma^z(r+\frac{e_z}{2})$ at the hinge, the domain wall point carries a spin-1/2 zero mode as Fig.~\ref{bosonhinge}. Such point defect on the hinge can be regarded as the corner of x-y domain wall hitting at the hinge. The global $Z_2$ symmetry for $\sigma$ spin guarantees the fluctuation and proliferation of domain wall at the hinge, while the $U^{sub}(1)$ and $\mathcal{T}$ protects the spin-1/2 zero mode(for $\tau$ spin) decorated inside the domain wall on the z-hinge. 

As the domain wall on the hinge carries a spinon degree of freedom, its condensate either breaks $U^{sub}(1)$ and $\mathcal{T}$ symmetry or result in a gapless state. Indeed this gapless hinge mode is described by $(1+1)D$ topological field theory with $O(4)_1$ WZW term\cite{bi2014anyon,xu2013wave,levin2012braiding,chen2013critical}.  
\begin{align} 
&\mathcal{L}_{\text{edge}}=\frac{1}{g}(\partial_{\mu} \vec{n})^2+\frac{2\pi}{\Omega^3} \int_0^1 du~ \epsilon^{ijkl}  n_i\partial_z n_j \partial_t n_k\partial_u n_l,\nonumber\\
 &\vec{n}(x,t,u=0)=(1,0,0,0),\quad
 \vec{n}(x,t,u=1)=\vec{n}(x,t).\nonumber\\
  &n_4=\langle \sigma_z \rangle, ~ n_{1}=\langle \tau_x \sigma_z \rangle ,~n_{2}=\langle \tau_y \sigma_z \rangle ,~n_{3}=\langle \tau_z \sigma_z \rangle 
\label{col}
\end{align}
The $O(4)$ WZW term implies that the domain wall of $\sigma_z$ carries a $(0+1)D$ $O(3)_1$ WZW term which exactly represents a spin-1/2 degree of freedom.
The global $Z_2$ and $\mathcal{T}$ symmetry act on the $O(4)$ vector boson as,
\begin{align}
&Z_2: \vec{n}(x,t) \rightarrow -\vec{n}(x,t);\nonumber\\
&\mathcal{T}: (n_1,n_2,n_3) \rightarrow (-n_1,-n_2,-n_3)
\end{align}
The subsystem U(1) symmetry rotates the 
hinge spin along the $\tau^z$ axis as,
\begin{align}
& e^{i \theta}=n_1+i n_2,\nonumber\\
& U^{sub}(1): e^{i \theta} \rightarrow e^{i \theta+\alpha}
\label{symm}
\end{align}

Obviously, the $O(4)$ WZW theory on the hinge is invariant under $U^{sub}(1) \times \mathcal{T} \times Z_2 $. These symmetries ensure none of the components in the O(4) vector can be polarized and the non-vanishing $O(4)_1$ WZW term always yields a gapless hinge akin to $(1+1)D$ $SU(2)_1$ CFT\cite{xu2013wave,xu2013nonperturbative}.

\subsection{LSM theorem perspective on surface}

Our previous analysis only demonstrates a specific way to gap out the side faces separated by a gapless z-hinge under symmetry protection. However, different from the justification we adopted for conventional SPT boundary, the stability of each symmetry enforced gapless hinge does not guarantee a nontrivial bulk topology. Despite the fact that the spatially separated hinges are stably gapless individually, they might still be hybridized and gapped via a surface phase transitions. It is still nebulous whether the hinge is ungappable or anomalous under any symmetry allowed surface reconstruction. To demonstrate the necessity of a gapless hinge, we here implement an LSM theorem argument to elucidate the ungappable nature of the hinge state.

The connection between ungappable boundary of an SPT surface and the absence of a featureless gapped insulator can be traced back to Ref.~\cite{else2019topological,jiang2019generalized,cheng2019fermionic,jian2018lieb,bultinck2018filling,hsieh2016all}. The generalized Lieb-Shultz-Matthis theorem states that some D-dim quantum many-body system with internal symmetry $G$ and lattice symmetry $S$ does not allow a featureless gapped phase with a unique ground state invariant under $G \times S$. In parallel with such lattice no-go theorem, an SPT boundary which is anomalous under $G$ and onsite $\tilde{S}$ symmetry does not support a featureless gapped surface state without breaking $G \times \tilde{S}$. Under ultraviolet regularizations, the lattice symmetry $S$ can be interpret into an internal symmetry $\tilde{S}$ and the low energy effective theory of $D$-dim SPT boundary with $G \times \tilde{S}$ is equivalent to the $D-1$-dim lattice model with $G \times S$ symmetry. This mapping has led to fruitful results: As an SPT boundary with $G \times S$ symmetry does not permit a trivial gapped boundary, the corresponding lattice spin systems cannot hold featureless gapped phase. Likewise, If a generalized LSM theorem implies the absence of featureless gapped ground state, one can extrapolate that the surface theory from dual-mapping is anomalous and thus must rely on a nontrivial bulk state.

Here we first explicate that our surface theory on xz and yz planes with $U^{sub}(1) \times \mathcal{T} \times Z_2 $ symmetry can be mapped into a 2D lattice model with $U^{sub}(1) \times \mathcal{T}$ and translation symmetry $T_z$.

In our current settings, each x-y domain wall corner hitting at the xz side face carries a spin-1/2 degree of freedom. Such domain wall corner on the side face can be defined as,
\begin{widetext}
\begin{align}
\sigma^z(r-\frac{e_x+e_z}{2}) \sigma^z(r-\frac{e_x-e_z}{2})\sigma^z(r+\frac{e_x+e_z}{2}) \sigma^z(r+\frac{e_x-e_z}{2})=-1
\label{def}
\end{align}
\end{widetext}

\begin{figure}[h]
  \centering
      \includegraphics[width=0.4\textwidth]{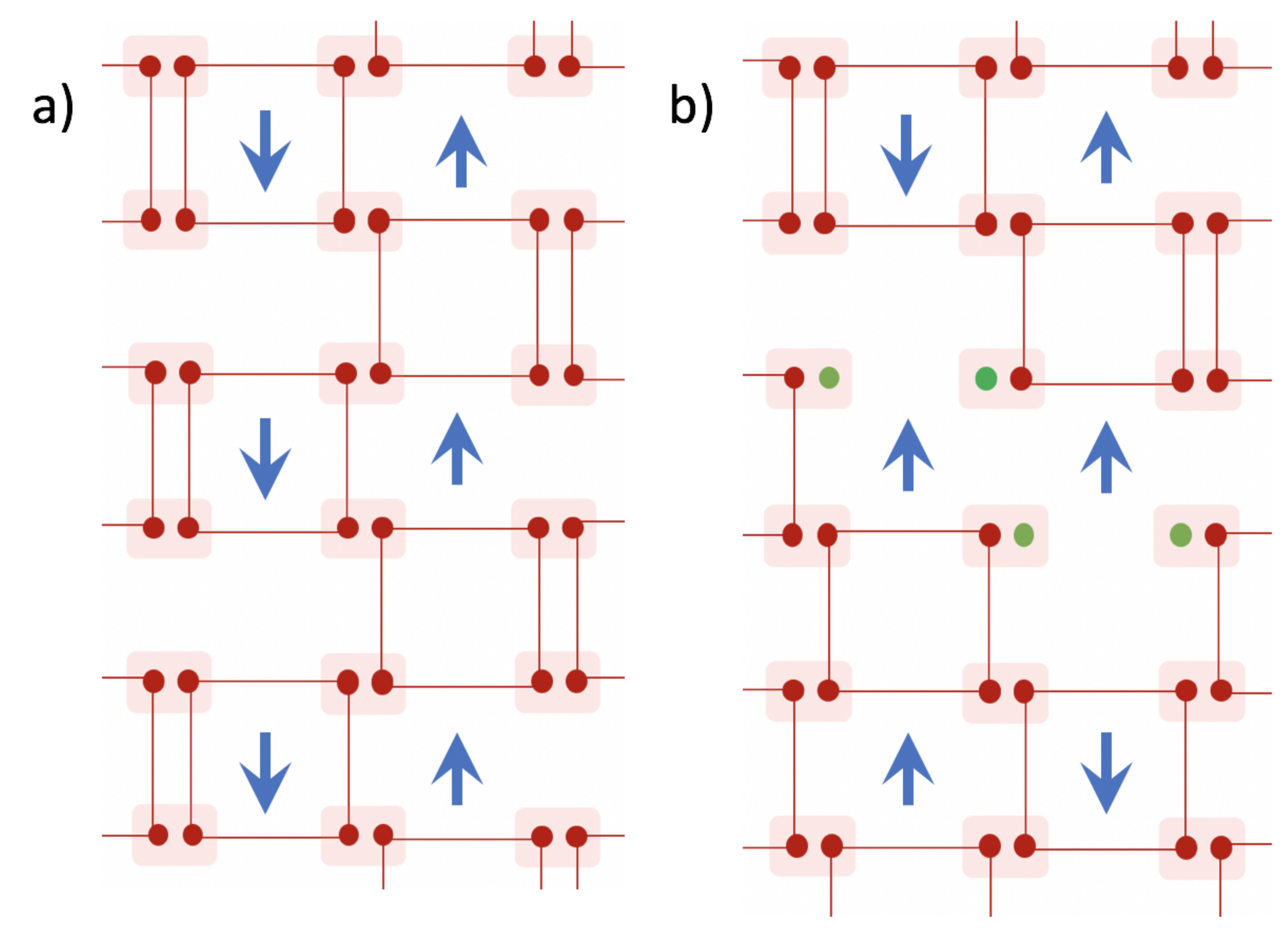}
  \caption{a) Square lattice with two spin-1/2 per unit cell. The red square refers to the valence plqauette state. The blue arrow characterize the $Z_2$ variable of the plaquette order parameter $Q$. b) When the plaquette order parameter $Q$ has a point defect, each defect carries an unpaired spin-1/2 (green dot).} 
  \label{sideb}
\end{figure}

Now we are about to map the surface theory into a 2D square lattice model with two(or even) spin-1/2 degree of freedom per site.
We first define a valence plaquette state as Fig.~\ref{sideb} where four spin-1/2 from the four corners of a plaquette are entangled as $|0101\rangle+|1010\rangle$. Such valence plaquette state preserves $U^{sub}(1) \times \mathcal{T}$ symmetry defined in Eq.~\ref{symm}.

Based on such valence plaquette configuration which breaks lattice translation, one can map the plaquette order into a $Z_2$ variable,
\begin{align}
Q(r)=(P(r+e_z/2)-P(r-e_z/2)), P \in 0,1
\end{align}
Where $P=1\,(0)$ corresponds to the valence plaquette occupancy (vacancy) on each square.  Q(r) is thus a $Z_2$ variable characterizing the valence plaquette order. 
In particular, such plaquette order is odd under translation symmetry along z-direction so the unit cell is doubled, 
\begin{align}
T_z: Q(r) \rightarrow Q(r+e_z)=-Q(r)
\end{align}
Such symmetry action resembles the global $Z_2$ symmetry acting on the $\sigma^z$ degree of freedom at the surface of HOSPT.
Further, If the plaquette order has a point defect as,
\begin{widetext}
\begin{align}
Q(r-\frac{e_x+2e_z}{2}) Q(r-\frac{e_x-2e_z}{2})Q(r+\frac{e_x+2e_z}{2}) Q(r+\frac{e_x-2e_z}{2})=-1
\label{def2}
\end{align}
\end{widetext}
Such point defect carries a spin-1/2 zero mode due to the unpaired spinon surrounded by the plaquette. This defect structure resembles Eq.~\ref{def} where the corner defect of four adjacent $Z_2$ variable carries a spin-1/2 mode. 
Based on these observations, we map the surface of HOSPT into a 2D lattice model. The primary degree of freedom on both side of the mapping involves a $\tau$ spin with $U^{sub}(1) \times \mathcal{T}$ symmetry. In addition, the 
valence plaquette ordering in 2D square lattice can be mapped into a $Z_2$ variable Q(r) which is odd under discrete translation $T_z$. Likewise, the $\sigma^z$ spin on the surface of HOSPT is odd under global $Z_2$. On both side of the mapping, the plaquette defect in Eq.~\ref{def2} and the $\sigma^z$ corner defect in Eq.~\ref{def} carry a spin-1/2 degree of freedom. Subsequently, the effective theory of 2D spin model with $U^{sub}(1) \times \mathcal{T}\times T_z$ on square lattice is equivalent to the HOSPT surface with $U^{sub}(1) \times \mathcal{T}\times Z_2$ symmetry. A no-go theorem for a featureless gapped ground state on the square lattice can be
extrapolated to the absence of symmetry invariant gapped HOSPT surface.

For square lattice with two spins per unit cell, one can construct a featureless gapped ground state which preserves $U^{sub}(1) \times \mathcal{T}$ and translation symmetry $T_z$.
To gain intuition, we first interpret the Pauli spin operators in terms of hardcore bosons:
\begin{align} 
\tau^x+i \tau^y=a^{\dagger},\tau^x-i \tau^y=a, \tau^z=a^{\dagger}a-1/2.
\end{align}
Each hardcore boson has a restricted onsite filling  $a^{\dagger}a=0,1$, and the states $|0 \rangle $ and $|1 \rangle = a^\dag |0 \rangle$ carry a $U(1)$ charge of $-1/2$ and $1/2$, respectively.  The boson creation operator $a^{\dagger}$ creates an $S_z = 1$  magnon excitation by changing $S_z\rightarrow S_z+1$.

In this language, the subsystem U(1) symmetry becomes a phase rotation for the bosons on each z-row at the side face,
\begin{align} 
& U^{sub}(1): a_j \rightarrow e^{i\theta } a_j,\;\; j \in \text{row}.
\end{align} Additionally, $\mathcal{T}$ acts as a particle-hole symmetry for the hardcore bosons,
\begin{align} 
\mathcal{T}:&|1 \rangle \rightarrow | 0 \rangle, |0 \rangle \rightarrow  - |1 \rangle \\
& a \rightarrow -a^{\dagger}, a^{\dagger} \rightarrow -a. 
\end{align}
For a $\mathcal{T}$ invariant ground state, the boson filling fraction is fixed as $\nu=2(S_z+1/2)=1$\footnote{There is a factor since there are two hardcore bosons(spin-1/2) per site.}.

Constructing a featureless gapped ground state is straightforward.
With two spin-1/2 per site, one can always pair the two spin per site into an onsite singlet as a unique gapped ground state which preserves the $U^{sub}(1) \times \mathcal{T}$ and translation symmetry. 

\begin{figure}[h]
  \centering
      \includegraphics[width=0.4\textwidth]{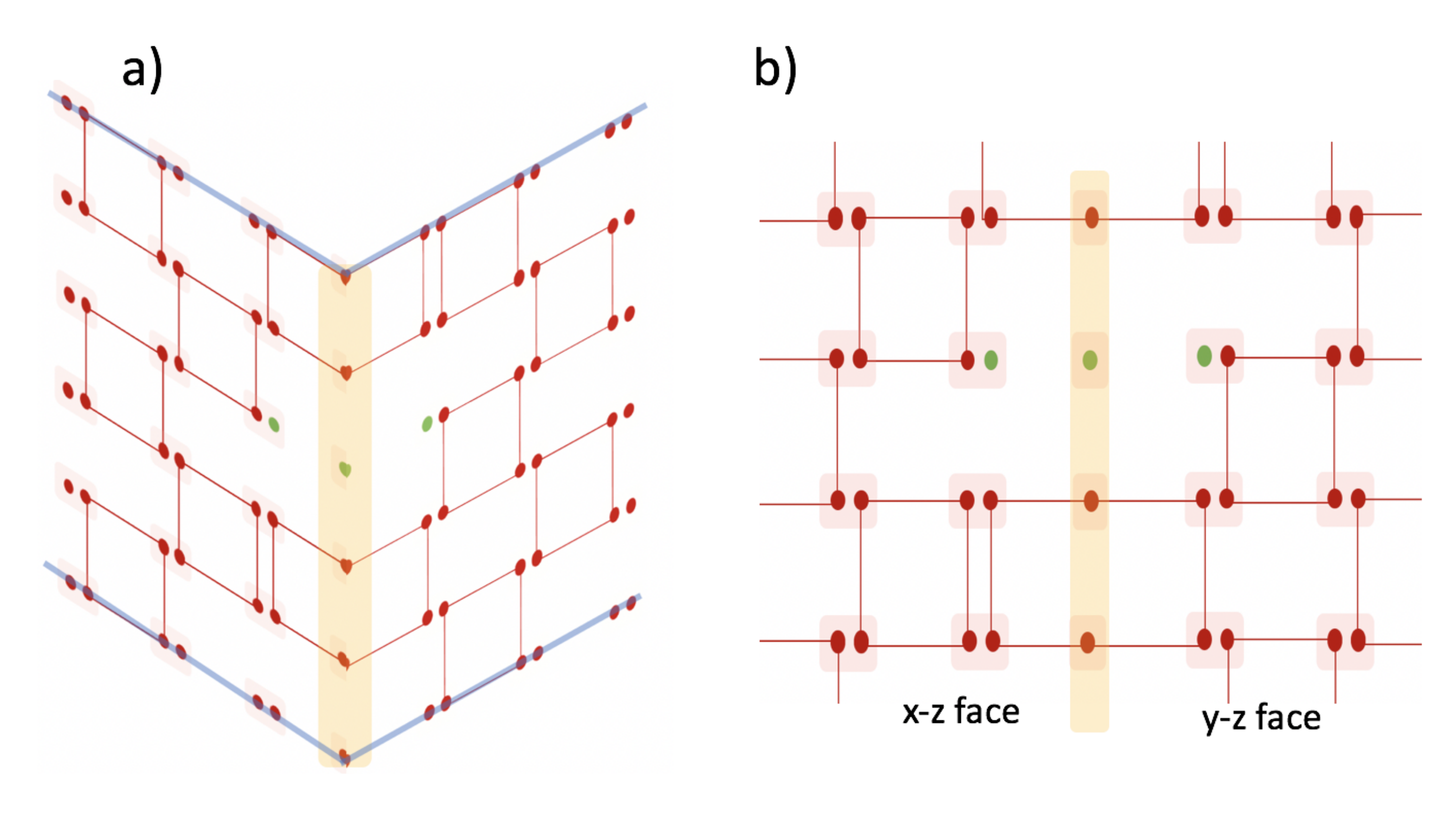}
  \caption{a) The hinge(yellow shaded area) of the HOSPT can be mapped into a spin 1/2 chain with odd number of spin per site. b) projected view of the z-hinge connecting two surfaces.} 
  \label{sideb2}
\end{figure}

However, when it comes with a hinge, the $Z_2$ domain wall of $\sigma^z$ carries an unpaired spin-1/2 on the hinge. Such hinge structure can be mapped into a spin chain with odd number of spin 1/2 per site as Fig.~\ref{sideb2}. The VBS order parameter plays the role as a $Z_2$ variable which is odd under translation $T_z$. Each VBS domain wall carries a spin-1/2 zero mode so the effective theory of such chain resembles the hinge of HOSPT. 

Based on this observation, the xz and yz surface can be mapped into a square lattice model with two(even) number of spin-1/2 per site. The hinge corresponds to a defect line along z-direction with odd number of spin-1/2 per site as Fig.~\ref{sideb2}. If the lattice system permit a featureless ground state with finite gap in the presence of translation and $U^{sub}(1) \times \mathcal{T}$, a necessary condition requires that each row along z-direction must carry even number of spin-1/2 per site. To be more explicit, we first assume there is a short-range entangled ground state which is invariant under lattice translation and subsystem U(1) symmetry on each z-row. Now we implement a $2\pi$ flux insertion on a specific z-row,

\begin{equation}
\begin{split}
U_{z,i} &= \exp\left(\frac{2\pi i}{L_z} \sum_{r \in i\text{-th row}} z \hat{n}_r\right),
\end{split}
\end{equation}
Such large gauge transformation creates a $2\pi$ flux for the $i$-th row along z direction. Due to the subsystem charge conservation, the global flux on each z-row are independent. For a featureless gapped ground state $|g\rangle$, such large gauge transformation leave the $|g\rangle$ state invariant. 

Meanwhile, as $|g\rangle$ is translation invariant under $T_z$, we can shift the ground state by one lattice spacing $a$ and leave it invariant. Combining these two operations together,

\begin{align}
&T_z U_{z,i} |g\rangle= e^{i2\pi \sum_{r \in i\text{-th row}} \hat{n}_r a/L_z } U_{z,i} T_z |g\rangle \nonumber\\
&=e^{i 2\pi \nu_i} U_{z,i}T_z |g\rangle
\end{align}
Here $\nu_i$ is the filling fraction of the spin-1/2 (hardcore boson) on the $i$-th row. When a specific z-row has odd number of spin-1/2 per site, we reach an obstruction that,
\begin{equation} 
T_z U_{z,i} |g\rangle= -  U_{z,i} T_z |g\rangle
\end{equation}
Which contradicts with the assumption that $|g\rangle$ is a unique ground state invariant under $U^{sub}(1)$ and translation. Thus, there is no featureless gapped ground state if there being a row with odd number of spin-1/2 per site. In the absence of hinge, each z-row has integer filling fraction so the side faces can be gapped. However, the hinge creates an extra row with half filling fraction whose ground state is either gapless or symmetry breaking.

\subsection{Top surfaces}
Now we consider the fate of the top surface on x-y plane. As the $S_z$ quantum number is conserved in each xz and yz plane, the top x-y surface has subsystem U(1) symmetry on each x and y row. When the open end of z-hinge defect touches the top surface, the AKLT chain decoration on the hinge leaves a spin-1/2 zero mode.  This top surface could be fully gapped by adding symmetry allowed perturbations. Based on the bulk Hamiltonian, the top surface emerges an unpaired $\tau$ spin when the adjacent $\sigma$ spin has a defect as,
\begin{widetext}
\begin{align}
\sigma^z(r+\frac{e_x+e_y}{2}) \sigma^z(r+\frac{e_x-e_y}{2})\sigma^z(r-\frac{e_x+e_y}{2}) \sigma^z(r-\frac{e_x-e_y}{2})=-1
\label{def}
\end{align}
\end{widetext}
As the $\sigma$ spin is in the paramagnetic phase, such point defect proliferates and the zero mode inside defect can give rise to a gapless surface.

To eliminate such zero mode carried by the defect, we add a surface interaction term on the top xy plane.
\begin{align} 
&H^1_{\sigma,\tau}= (1 +\sigma^z(r))  |\psi_{r} \rangle  \langle \psi_{r}|
\label{ps}
\end{align}
Such interaction decorates each $\sigma^z=-1$ on the top surface with a four $\tau$ spin entangled state $|\psi \rangle_{r}=|0101\rangle+|1010\rangle$ living at the corners of the plaquette surrounding $\sigma^z$. Based on such surface decoration structure, the defect in Eq.~\ref{def} contains even number of spin-1/2 zero mode which can be gapped via onsite interaction. In addition, the hinges along the x or y direction is also gapped so the gapless degree of freedom is only localized at the z-hinge.

\section{Fermionic HOSPT with gapless hinge mode}

Current interest in higher-order topology is driven primarily by the material realization of fermion band theories\cite{slager2013space,benalcazar2017quantized,schindler2018higher,PhysRevB.97.035138,schindler2017higher,po2017symmetry,2018arXiv180502598T}. This raises the question of whether HOSPT protected by subsystem symmetry can be realized in interacting fermion systems. As we had highlighted in previous discussions, the subsystem symmetric HOSPT states do not have non-interacting counterparts as any fermion bilinear term inevitably breaks subsystem symmetry. Thus, most prominent constructions for fermion HOSPT cannot be readily translated to interacting systems, which require a fundamentally different approach. 
In the process, we will uncover an alternative route by implement the decorated hinge-wall picture to fermion systems. This construction can be regarded as the fermionic version of the HOSPT we developed in Section III by replacing the $\tau$ spins decorated on the hinge-wall with Majorana fermions.

To begin let us look into a fermion model on BCC lattice in Fig.~\ref{fermion}. The cube center contains a single Ising spin $\sigma$(blue dot) while the cube corner carries eight Majoranas(red dots).
\begin{figure}[h]
  \centering
      \includegraphics[width=0.4\textwidth]{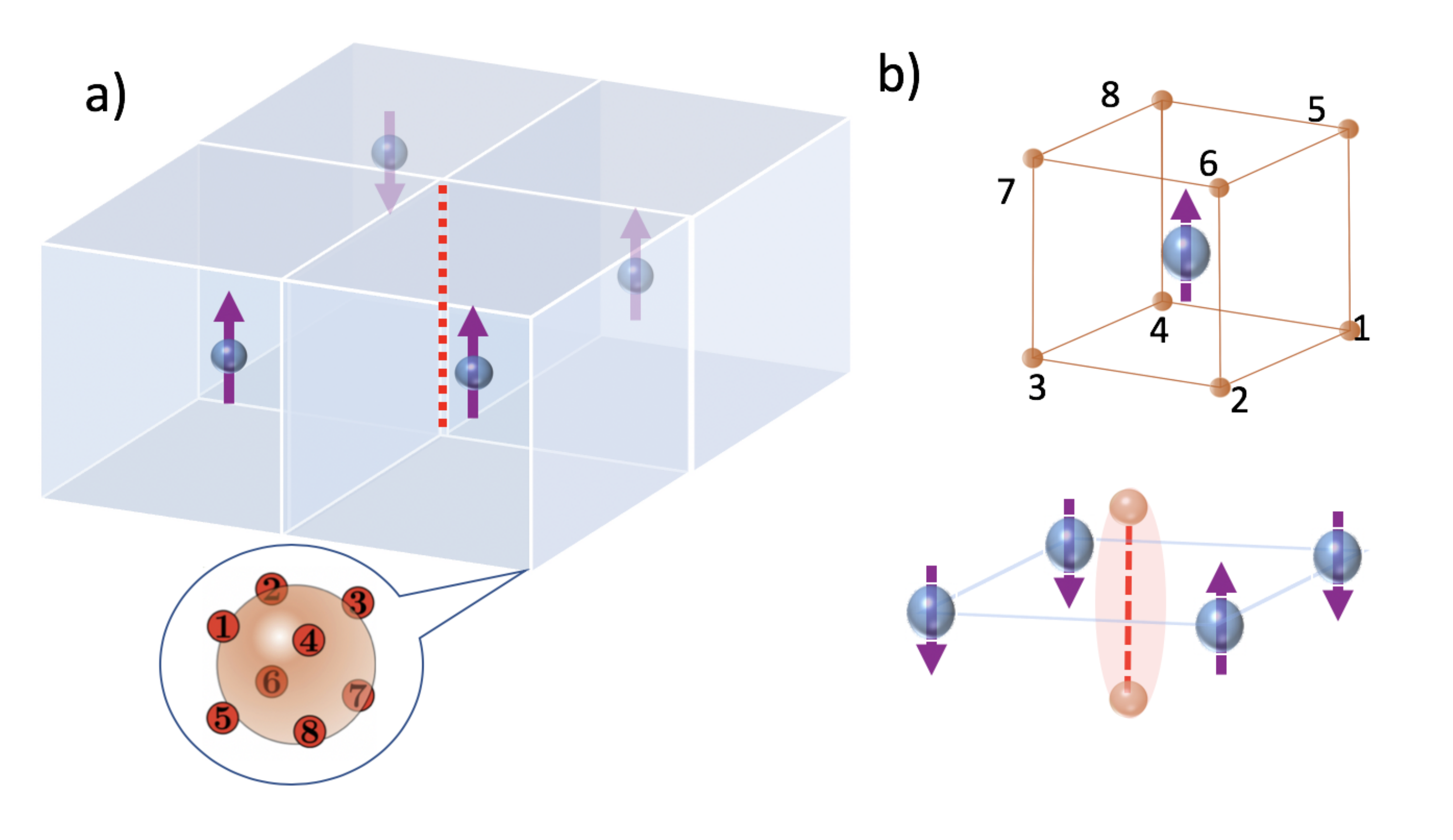}
  \caption{a) The cube center contains an Ising spin and the cube corner carries eight Majoranas. b) The eight Majorana at the corner of the cube is projected into a unique state if the center spin is polarized in up direction.  If the four $\sigma$ spins adjacent to the z-hinge contain odd number of $\sigma^z=-1$, we hybridize the two Majoranas $i\eta(r)\eta(r+e_z)$ between the z-link. } 
  \label{fermion}
\end{figure}
The spin $\sigma$ living at the cube center is placed into a $Z_2$ paramagnetic phase,
\begin{align} 
&H_{\sigma}=-\sum_i \sigma^x_i
\end{align}
Such Hamiltonian has a global $Z_2$ symmetry generated by $\sigma^x$. In the paramagnetic phase, the $\sigma$ spin are polarized in the x-direction. If we choose the $\sigma^z$ basis, the ground state wave function of such paramagnet is a coherent superposition of all domain wall membranes on the cubic lattice. 

The interactions among the Majoranas from cube corners depend on the adjacent $\sigma$ spin structure. In particular, for each cube center with $\sigma^z=-1$, we project the eight Majoranas from the eight corners of each cube into a unique ground state in the same way as Eq.~\ref{four}-\ref{four2}. As is elucidated Section II, such interaction preserves the subsystem fermion parity symmetry on both xz and yz plane. Based on the discussion in Section II, such decoration structure embellishes the $\sigma^z$ domain wall corner\footnote{The domain wall corner is the intersection point between three domain walls from xy,xz,yz planes} with a Majorana zero mode.

In the meantime, when the four $\sigma$ spins adjacent to each z-hinge contain odd number of $\sigma^z=-1$ as Fig.~\ref{fermion}, we decorate the z-hinge by creating a two Majorana entangled pair $i\eta(r)\eta(r+e_z)$ on the z-link. 
Such `hinge defect' can be regarded as the intersection line between two domain walls from the yz and xz plane. The Majorana hybridization embellishes each `z-hinge defect' with a Kitaev chain which still preserves subsystem fermion parity on xz and yz planes. The open end of the z-hinge defect contains a free Majorana zero mode. 

This Hamiltonian we construct here has a simple ground state wave function. As the $\sigma$ spin living at the cube center is in the paramagnetic phase, its ground state can be expressed in terms of superposition of all close domain wall configurations in the bulk. Meanwhile, due to the Majorana decoration, the domain wall on each x-y plane contains a 2d HOSPT with Majorana corner mode while the z-hinge defect is decorated with a Kitaev chain. As the z-hinge defect is nothing but the domain wall intersection line between xz and yz plane, any open end of the z-hinge defect should be connected to the domain wall corner of the xy plane.  As a result, the ground state can be viewed as a hinge-wall condensate with each z-hinge connecting a domain wall corner from the xy plane.
The Majorana zero mode contributed from the corner of x-y domain wall can hybridize with the Majorana zero mode contributed from the end of the z-hinge defect. Subsequently, the decorated hinge wall condensate generates a gapped phase with a unique ground state.

When it comes to the surface on yz and xz planes, the corner of the x-y domain wall hitting the side faces on yz(or xz) plane is no longer connected by a z-hinge. Such open x-y domain wall corner at the boundary carries an unpaired Majorana zero mode. Follow the procedure in section III, we can construct a symmetry preserving surface perturbations which fully gap out the degree of freedom on the side faces.
By decorating the z-directional domain wall line on each each xz(yz) surface with a Kitaev chain, the side face is fully gapped.

We now consider the degrees of freedom on the z-hinge. Based on our construction, if there being a domain wall between $\sigma^z(r-\frac{e_z}{2}),\sigma^z(r+\frac{e_z}{2})$ at the hinge, the domain wall point carries a Majorana zero mode. The global $Z_2$ symmetry for $\sigma$ spin guarantees the fluctuation and proliferation of domain wall, while the subsystem fermion parity protects the Majorana zero mode decorated inside.

Such hinge structure can be mapped into 1D Majorana chain with odd number of Majorana per site. One can further define a bond order parameter,
\begin{equation} 
Q= \text{sgn}( \langle i\eta_{2i}\eta_{2i+1}-i\eta_{2i-1}\eta_{2i}\rangle)
\end{equation}
which is odd under lattice translation $T_z$. We can map this $Z_2$ bond order to the Ising variable $\sigma^z$ on the hinge of HOSPT and the lattice translation $T_z$ plays a role as the onsite $Z_2$ symmetry at the HOSPT hinge.
 As the domain wall of $Q$ carries a Majorana zero mode, the effective theory of such Majorana chain resembles the hinge of our HOSPT. 
 For 1D Majorana chain with odd Majorana per site and a conserved fermion parity, the LSM theorem\cite{hsieh2016all} predicts the absence of any featureless gapped state. Such no-go theorem can be extrapolated to the HOSPT hinge whose low energy effective theory is either gapless or breaks fermion parity and $Z_2$ symmetry. In conclusion, the present model seeks a fermionic HOSPT with gapless hinge protected by $Z_2$ and subsystem fermion parity.

\section{Outlook}

In this work we proposed several interaction enabled higher-order topological phases with subsystem symmetry protection.
We identified a series of solvable models that engender gapless hinges or corner states protected by subsystem symmetry and global symmetries. Remarkably, the ungappable nature of the HOSPT hinges can be connected to the `LSM-type' no-go theorem in lower dimensional lattice system which does not permits a featureless gapped ground state.

We expect the main logic and method used in this paper can be generalized to other subsystem symmetries and thus predicts a complete classification of HOSPT enabled by subsystem symmetry. Such exploration also shed light on the search for symmetry enriched fracton phases which can potentially host abundant and fascinating phenomenology. 

\textit{Acknowledgments}: YY is supported by PCTS Fellowship at Princeton University. This work was initiated at Aspen Center for Physics, which is supported by National Science Foundation grant PHY-1607611.

 \end{document}